\documentclass[twocolumn,fleqn]{svjour2}    

\smartqed  
\usepackage{graphicx,color}

\begin{document}

\title{A novel internal waves generator}

\author{L.~Gostiaux  \and H.~Didelle \and S.~Mercier \and T.~Dauxois}

\institute{L. Gostiaux, T. Dauxois \at
Laboratoire de Physique,
UMR-CNRS 5672, ENS Lyon, 46
  All\'{e}e d'Italie, 69364 Lyon c\'{e}dex 07, France\\   Email: lgostiau@ens-lyon.fr,Thierry.Dauxois@ens-lyon.fr    \\
           \and H.~Didelle, S.~Mercier \at
Laboratoire des \'Ecoulements G\'eophysiques et Industriels (LEGI),
UMR 5519 CNRS-UJF-INPG, 21 rue des Martyrs, 38000 Grenoble, France}

\date{Received: date / Accepted: date}

\maketitle

\begin{abstract}
We present a new kind of generator of internal waves which has been
designed for three purposes. First, the oscillating boundary
conditions force the fluid particles to travel in the preferred
direction of the wave ray, hence reducing the mixing due to forcing.
Secondly, only one ray tube is produced so that all of the energy is
in the beam of interest. Thirdly, temporal and spatial frequency
studies emphasize the high quality for temporal and spatial
monochromaticity of the emitted beam. The greatest strength of this
technique is therefore the ability to produce a large monochromatic
and unidirectional beam.
 \keywords{Internal waves \and Stratified fluids}
\end{abstract}

\section{Introduction}\label{intro}

Stratified fluids are nowadays a field of paramount importance in
fluid mechanics and are carefully studied theoretically,
experimentally and numerically by a large and always increasing
scientific community. The underlying reason is, of course, the
identification of many crucial phenomena appearing in the oceanic
and atmospheric environments, where the stratification cannot be
overlooked. The possibility to reproduce in a laboratory experiment
conditions close to  theoretical hypotheses is, of course, a very
important way to validate theoretical ideas  developed in the last
decades. Indeed, if one will always rely on {\em in situ}
observations for the final decisive tests, new fundamental ideas are
usually derived within an oversimplified framework that cannot be
reproduced in a real environment. Among many difficulties, let us
mention just a few. The linear stratification hypothesis is almost
always used in theoretical works despite the complicated oceanic
or atmospheric stratification (Thorpe 2005)
. Internal waves are also often considered as monochromatic plane
waves, while oceanic measurements usually report a beam-like
structure, with a width smaller than a
wavelength (Lam et al. 2004, Gerkema et al. 2004)
. The paradox of internal waves impinging onto topography (Dauxois
and Young 1999) 
 was solved for a linear oblique slope
which can only be a poor description of a real seamount (Eriksen
1998)
. It is not satisfactory to test this last theoretical example by
considering a {\em beam} impinging onto a {\em curved} topography in
a fluid with a {\em real and complicated} stratification
(exponential, thermocline, mixed layer,...). Indeed, possible
disagreements between theoretical and experimental results might
always be attributed to one of these three hypothesis which are not
fulfilled in the experimental test. Controlled experiments in a
simplified framework are therefore of paramount importance.

In a laboratory experiment, one can identify three main
difficulties: the preparation of a stratified fluid with a
controlled {\em stratification}, the {\em generation} and the {\em
visualization} of internal waves propagating within this fluid. The
first difficulty was rapidly solved thanks to the well-known double
bucket method and its improved versions (Oster 1965, Hill 2002,
Benielli and Sommeria
1998)
. The visualization difficulty has been nicely solved in the last
decade by two different techniques: the Schlieren method, on the one
hand, later improved into the modern quantitative
synthetic Schlieren method (Dalziel et al 2000)
, gives time resolved density fields; on the other hand, the
particle image velocimetry (PIV) technique provides time resolved
velocity fields
(Fincham and Delerce 2000)
. However, the generation of monochromatic plane waves remains a
crucial problem from the experimental point of view. We propose in
this article a new experimental set-up  that is able to provide a
single monochromatic and unidirectional large beam.

The paper is organized as follows. In section~\ref{shortreview}, we
briefly review existing generation methods, with a special emphasis
on their limitations. In Sec.~\ref{newgenerator}, we present the new
internal waves generator. In the following
section~\ref{analysisofinternalwaves}, we analyze the internal waves
which are generated with a particular emphasis on its temporal, and
spatial properties. Section~\ref{conclusion} concludes the paper.

\section{Review of existing generation methods}
\label{shortreview}

\subsection{Important properties of internal gravity waves}

Internal waves in a continuously stratified fluid are shear waves
that propagate with a phase velocity perpendicular to their group
velocity. The density stratification allows the propagation of such
waves by generating a restoring force. This latter is characterized
by the  Brunt-V\"ais\"al\"a frequency $N$, which corresponds to the
vertical oscillation frequency of water masses in a local density
gradient. It is defined as $N^2=-(g/\rho)(\mbox{d}\rho/\mbox{d} z)$
where $g$ is the gravity, $\rho$ the density and $z$ the vertical
coordinate. The frequency $N$ will be kept constant in the remainder
of the paper.

In such a stratified fluid, it is well-known that the dispersion
relation for linear internal gravity waves is
\begin{equation}
\omega=N\sin\theta,\label{disprelat}
\end{equation}
where $\omega$ is the temporal frequency and $\theta$ the angle of
the wave vector with the vertical direction. This dispersion
relation emphasizes that for a fixed frequency, the direction in
which the shear propagates is fixed and is denoted~$\sigma$. This
direction is also found to be orthogonal to the direction of
propagation of the energy, denoted~$s$.


The study of internal waves generation must therefore take into
account these very unusual and restrictive geometrical conditions.
Moreover, it is very important to notice that the spatial wavelength
of the wave does not appear in the dispersion
relation~(\ref{disprelat}). There is no mechanism of wavelength
selection other than the physical boundary conditions imposed by the
forcing technique. This last point will be of first importance in
the following.

\subsection{Oscillating bodies}

\begin{figure}[!h]\begin{tabular}{cccc}
\includegraphics[width=0.375\columnwidth]{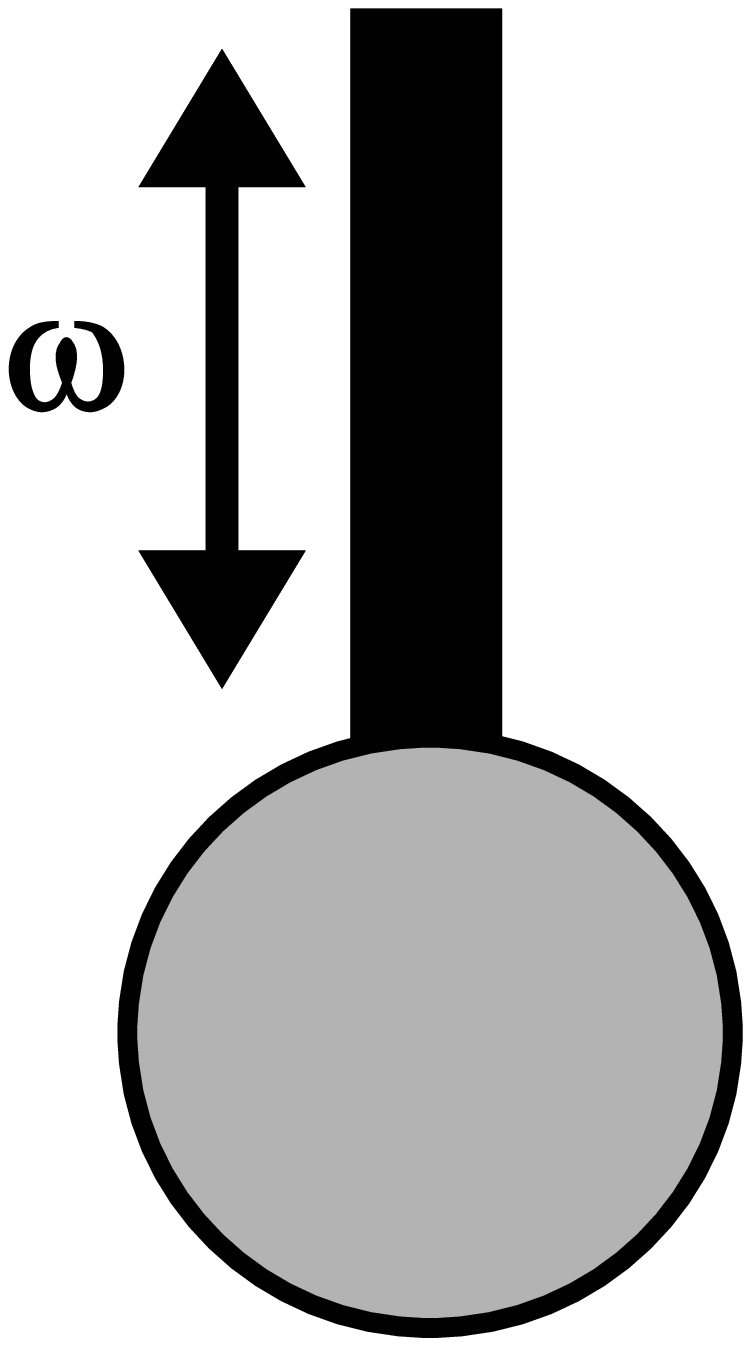} &{\bf a)}& \includegraphics[width=0.375\columnwidth]{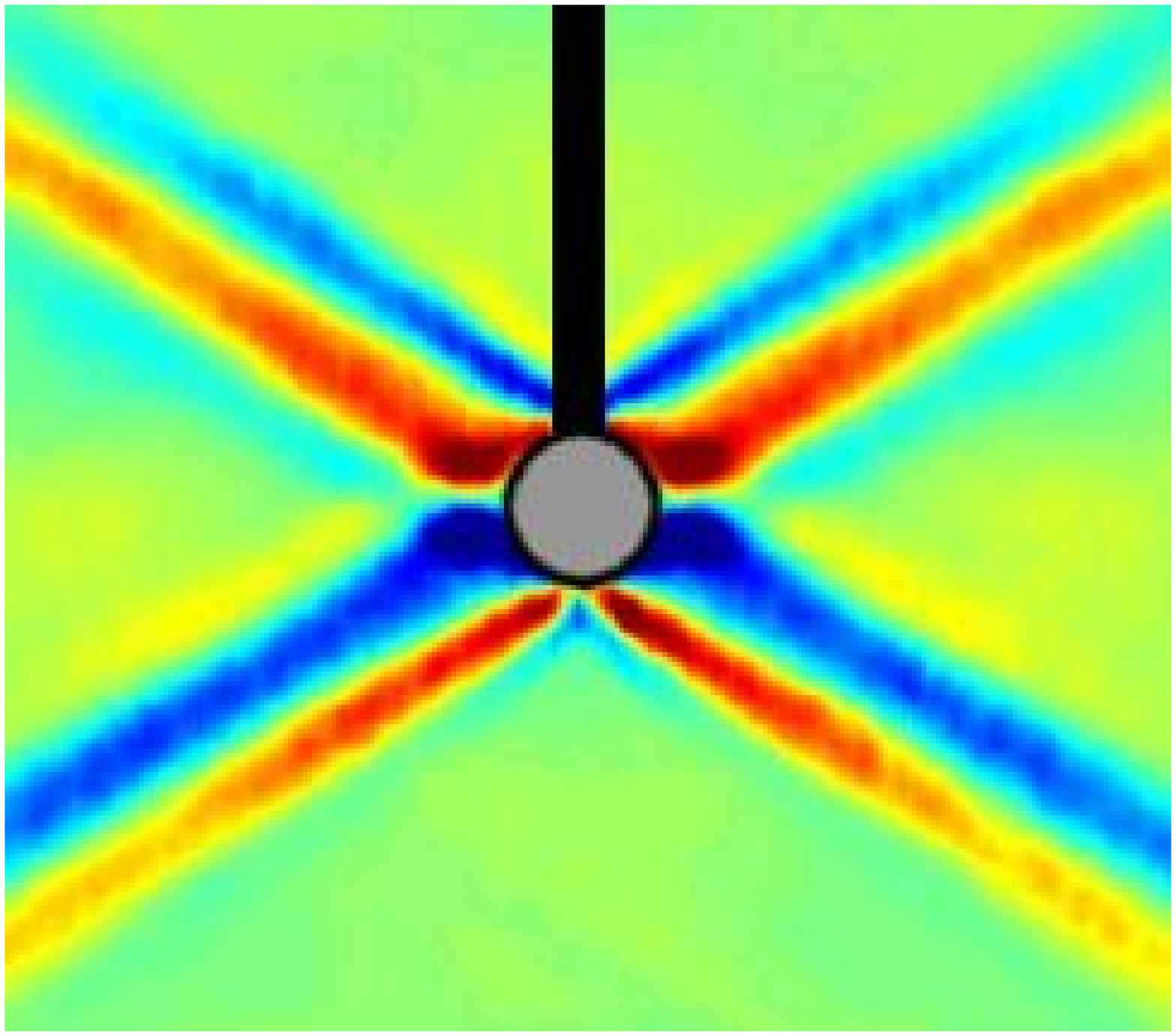}&{\bf b)} \\
\includegraphics[width=0.375\columnwidth]{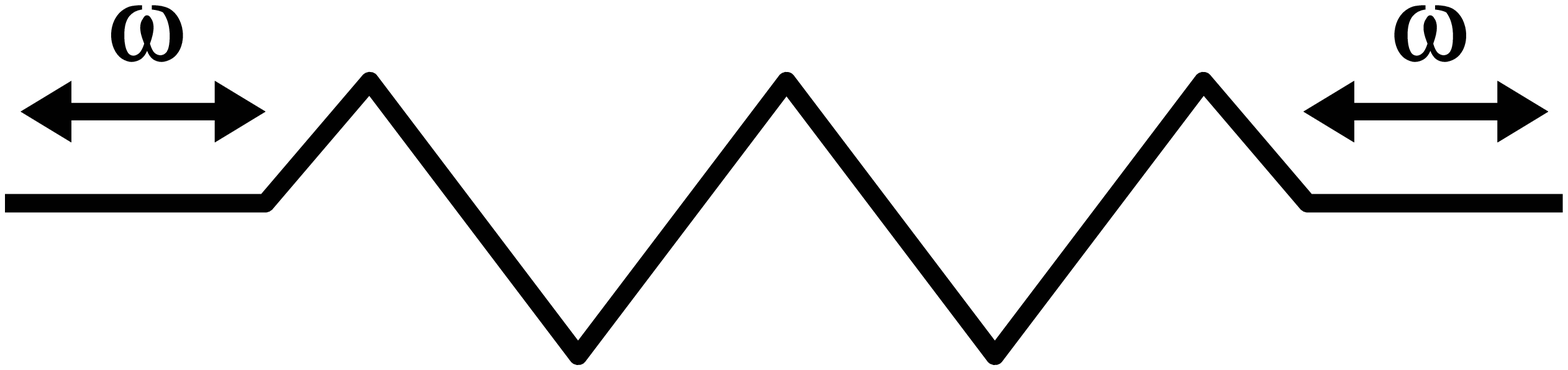} &{\bf c)}& \includegraphics[width=0.375\columnwidth]{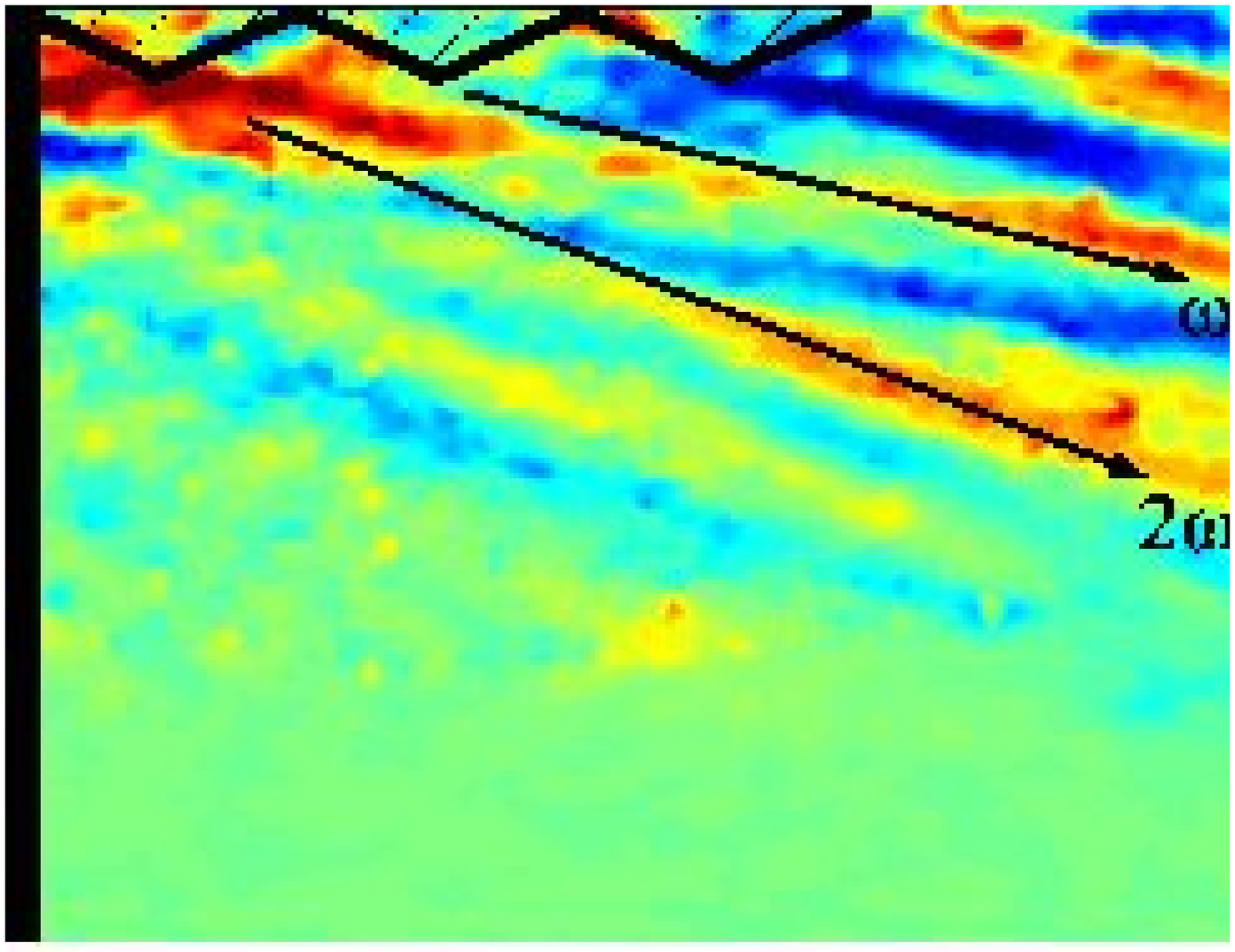}&{\bf d)} \\
\includegraphics[width=0.375\columnwidth]{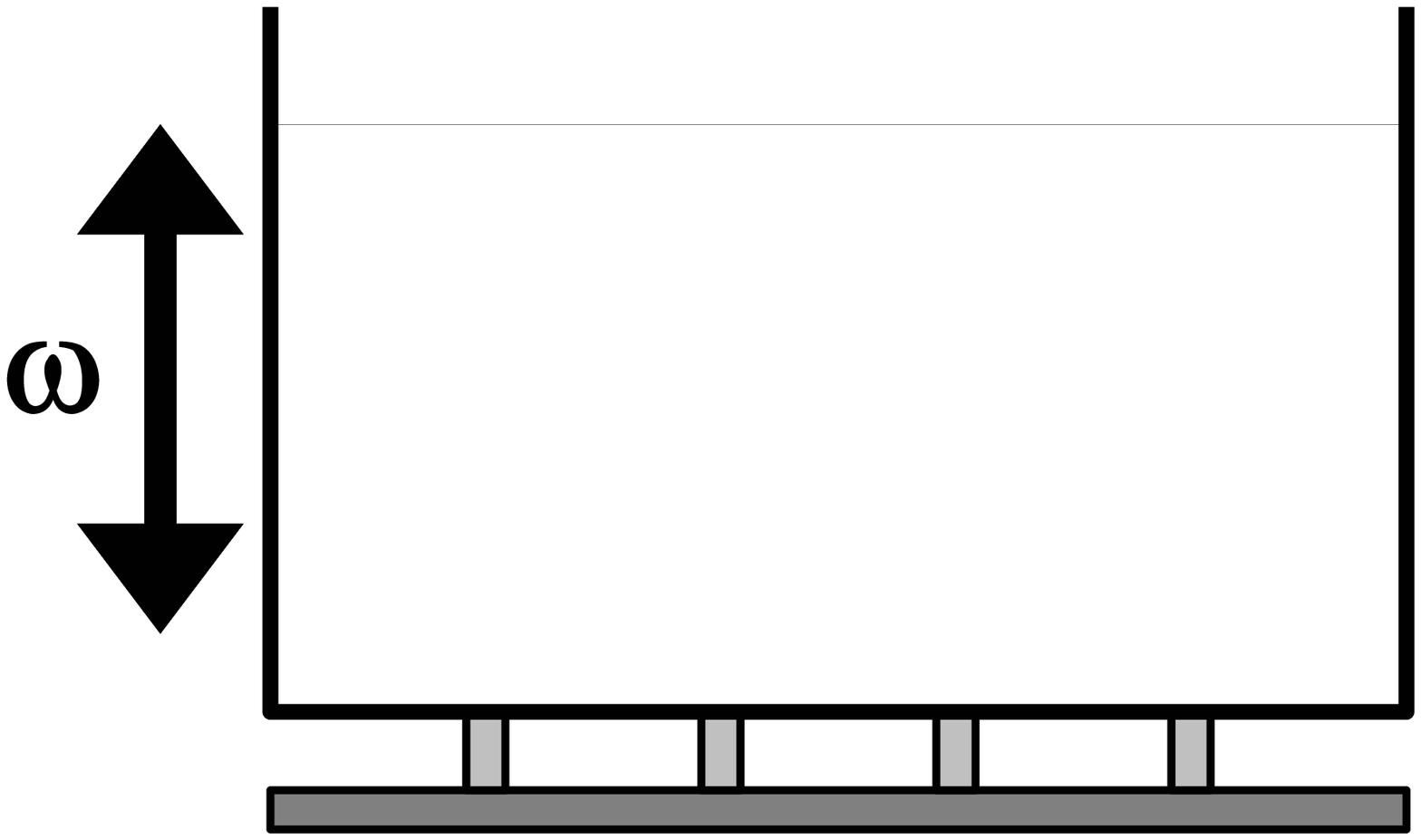} &{\bf e)}& \includegraphics[width=0.375\columnwidth]{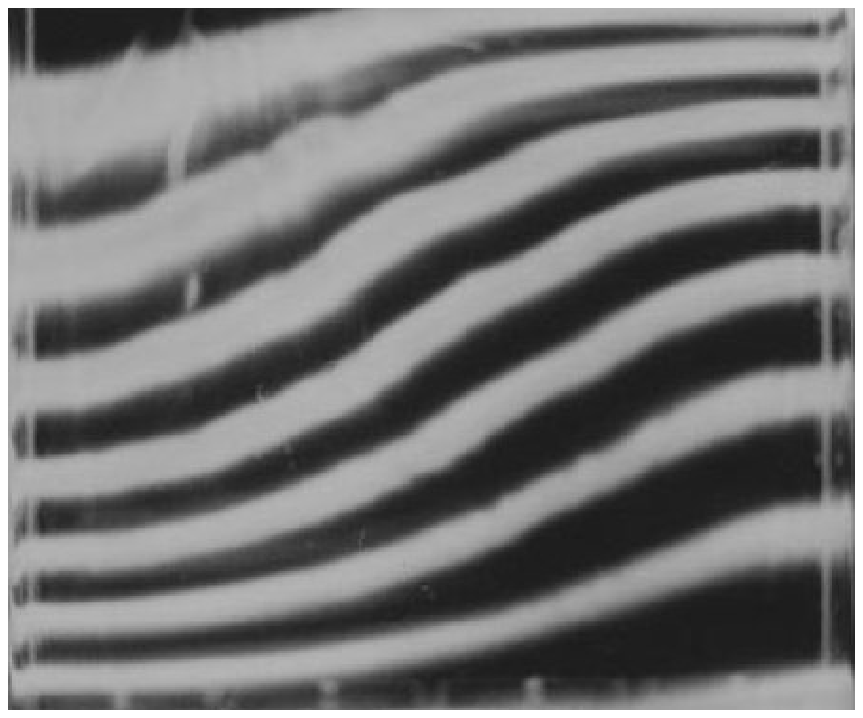}&{\bf f)} \\
\end{tabular}
\caption{(Color online) Principles and internal waves patterns for
the three previous internal gravity wave generators. Panel (a)
presents the sketch of a vertically oscillating cylinder. Panel (b)
shows the resulting four rays as captured with a synthetic Schlieren
technique. Panel (c) presents the paddle-like generator, while panel
(d) shows a typical instantaneous PIV measurement. Note harmonic
waves propagating in different directions.  Panel (e) presents the
parametric excitation principle while panel (f) shows the resulting
internal wave pattern (Benielli 1995).
} \label{fig:schem_emission}
\end{figure}

The simplest experiments on internal waves emission use oscillating
bodies as radiation sources and, more precisely, in a
two-dimensional set-up, a vertically oscillating cylinder (see the
sketch in Fig.~\ref{fig:schem_emission}(a)). The first experiment
was
performed by G\"ortler (1943) 
 in a two-dimensional setup. Nonetheless, as this experiment is often forgotten, one usually refers to
the seminal experiment by Mowbray and Rarity (1967). 
 This is nowadays a simple laboratory set-up, often used in
laboratory tutorials at the undergraduate level.  In the simplified
two-dimensional set-up, these generators emit four beams of internal
waves, making a constant angle with the horizontal. A typical
resulting pattern is presented in Fig.~\ref{fig:schem_emission}(b).
Recent three dimensional experiments with an oscillating sphere have
reported striking double cones (Peacock and Weidman 2005).

As the wavelength does not enter the dispersion
relation~(\ref{disprelat}) of internal waves, it is absolutely non
trivial, {\em a priori}, to predict the spatial structure of the
beam. This question has been at the origin of several theoretical
studies (Thomas and Stevenson 1972, Hurley 1997, Hurley and Keady
1997, Hurley and Hood 2001, Voisin
2003) 
 which have proposed a nontrivial
scaling of the beam spatial structure by the size of the object, the
viscosity of the fluid and the emission frequency. In all cases, the
beam envelope does not contain a full wavelength. Therefore, it is
difficult to use these sources to study spatial properties of
internal waves.

\subsection{Paddle-like generators}

It has also been  proposed to generate directly a shear motion in
the stratified fluid by using a multi-bladed folding paddle (McEwan
1973, Cacchione and Wunsch 1973, Teoh et al. 1997, Silva et al.
1997, Ivey et al. 2000, Gostiaux et al. 2006)
: see the sketch in Fig.~\ref{fig:schem_emission}(c). The flow
generated by such a device corresponds to the superposition of two
shear waves propagating in opposite directions along the zig-zag
paddle, each one being independently solution of the internal waves'
equations. Such a set-up thus generates  two beams propagating
symmetrically leftwards and rightwards (resp. upwards and downwards)
when the paddle is set horizontally (resp.
vertically) as in Teoh et al. (1997). 
 Locating a vertical wall close to the paddle (see vertical black
thick line in Fig.~\ref{fig:schem_emission}(d)) allows to avoid a
second beam: moreover, because of the reflection of the leftward
beam on this wall, one may increase artificially the width of the
rightward beam.

The energy transmission is proportional to the sine of the angle
between the paddle and the along-direction of the beam. To gain in
amplitude, one might therefore propose to tilt the paddle
orthogonally to the desired beam. One realizes immediately, however,
that the generated shear is  no more the  superposition of two
internal wave beams. Consequently it will excite strong nonlinearities.
Intermediate position between the horizontal and this inclination
can be proposed,  this method, however, still generates higher
harmonics. As shown in Fig.~\ref{fig:schem_emission}(d) (see
Gostiaux et al. (2006) 
 for additional details),
the main part of the energy is transmitted to a beam of internal
wave of frequency $\omega$, while higher harmonics $n\omega$ are
also excited by the oscillating paddle. Their frequencies being
higher, their propagation angles are also larger (see
Fig.~\ref{fig:schem_emission}(d)) and a screen can therefore be
located appropriately to remove most harmonics (Gostiaux et al.
2006)
. Fortunately, in Ivey (2000)
, this effect was avoided since the excitation frequency $\omega$
was larger than half of the Brunt-V\"ais\"al\"a. In conclusion, as
avoiding harmonics generation is of primary importance for energy
fluxes measurements and comparisons with theoretical predictions,
especially for low frequency ratios $\omega/N$, such a method is not
ideal.

\subsection{Tidal-like excitation}

Finally, a third way of generating internal waves has been proposed
and realized in a form of a global excitation of internal waves. A
vertical periodic motion of the fluid container (see the sketch in
Fig.~\ref{fig:schem_emission}(e)) acts as a perturbation of the
gravity force on the whole fluid. Tidal-like excitations by
barotropic forcing (Ivey 89, Gostiaux
2006) 
 also excite globally the fluid. Resonant modes are selected and can be focused on internal waves
attractors. This excitation is highly dependent of the size and
shape of the fluid domain, and relies on resonances of the enclosed
geometry itself (Maas et al. 1997) 
 as shown in
Fig.~\ref{fig:schem_emission}(f).

The first very limitation is that  oscillating vertically a large
tank is, of course, very difficult and cannot be easily performed.
In addition, the whole stratified fluid is thus excited so that it
is not possible to discern how internal waves, generated in one part
of the tank, might encounter independently a topography, or another
internal wave, in another part of the tank.

\section{The new solution for internal waves generation}
\label{newgenerator}

As briefly motivated in the previous section, a new principle for
internal wave  generation has to be proposed, if a {\em
monochromatic plane wave} is necessary for laboratory experiments.
The key idea is to create physical boundary conditions that
effectively propagate within the fluid interior, instead of a
stationary forcing.

\subsection{Mechanism of the generator}

\begin{figure}[ht]
\hskip
0.7cm\includegraphics[width=0.25\columnwidth]{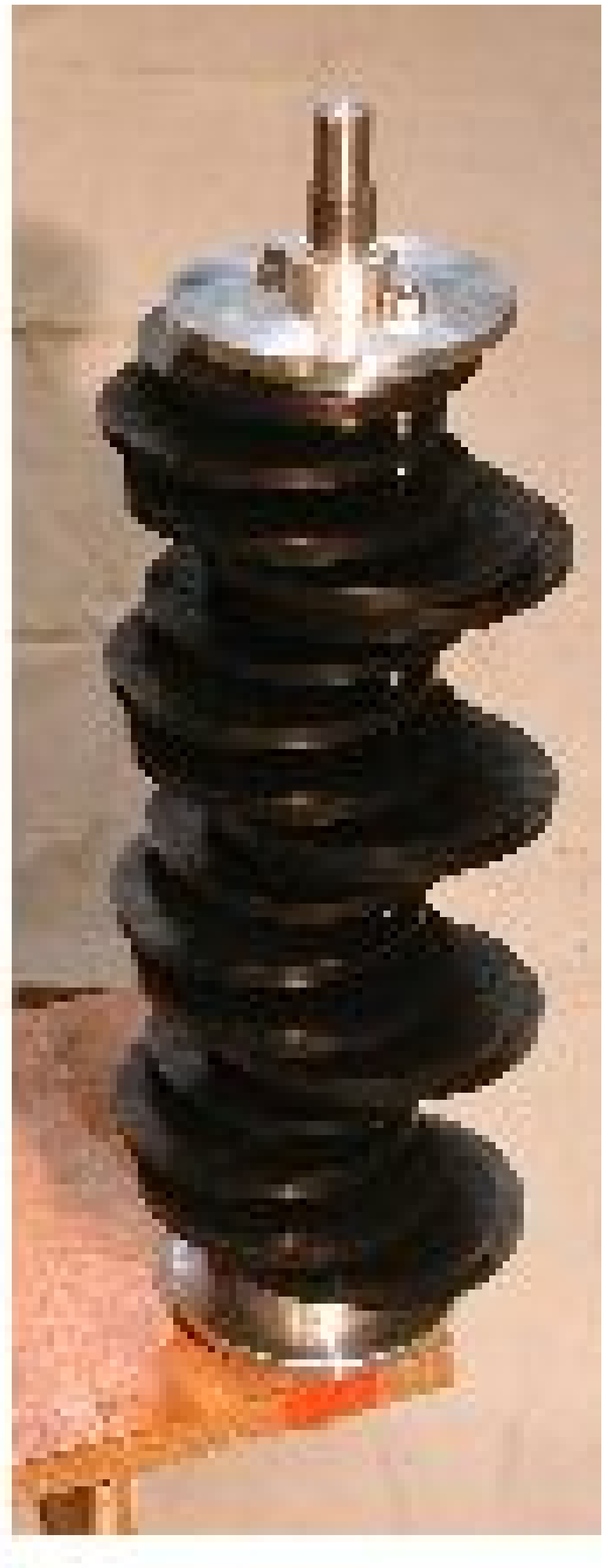}\hskip1truecm
\includegraphics[width=0.55\columnwidth]{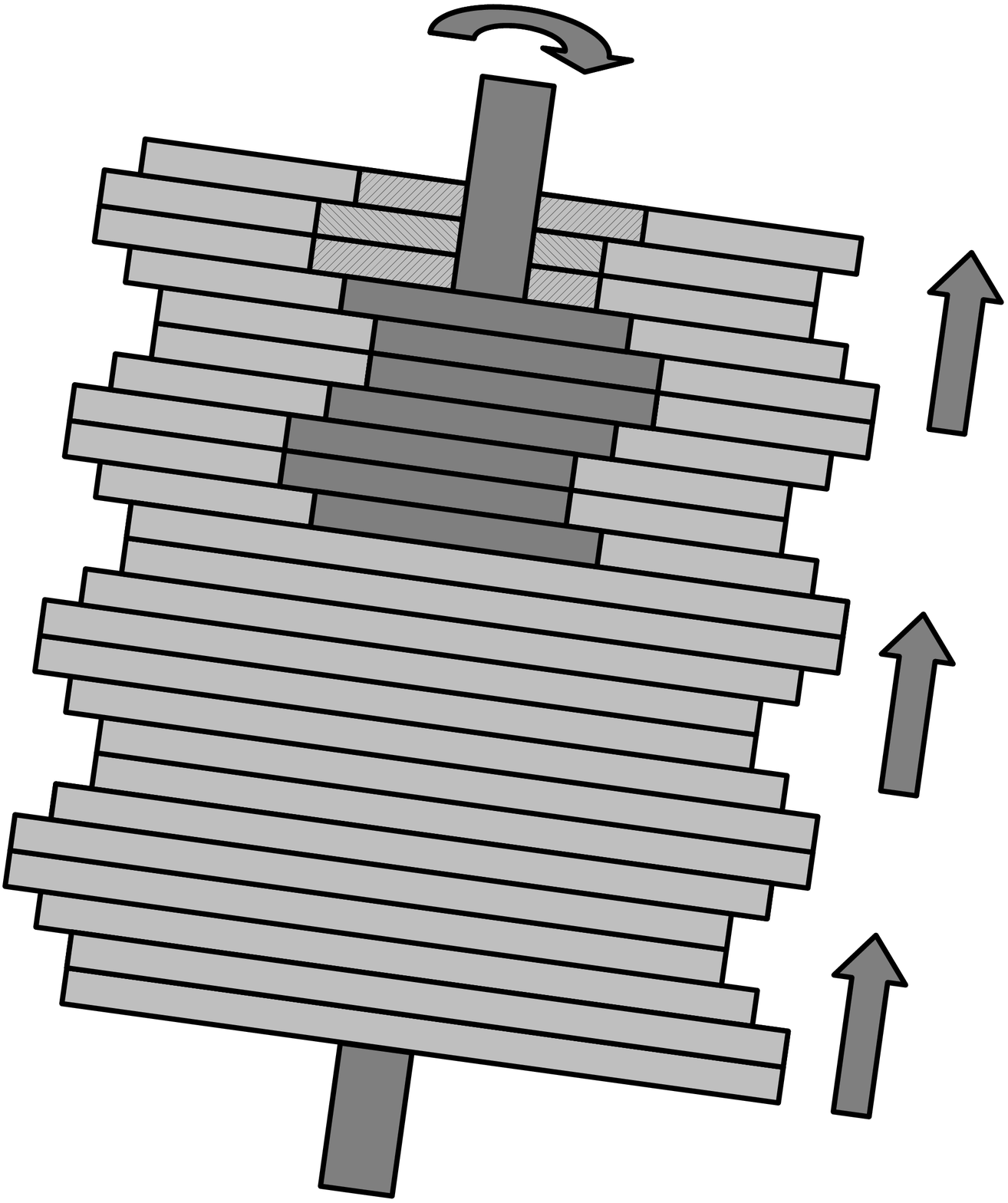}
\caption{Internal wave generator mechanism. The left panel presents
one of the two identical camshafts, while the right one shows the
cross section of the generator. One clearly identify the plates
(light grey) and one of the camshafts (dark grey).}
\label{fig:schem_excit}
\end{figure}

\begin{figure}[!h]
\includegraphics[width=\columnwidth]{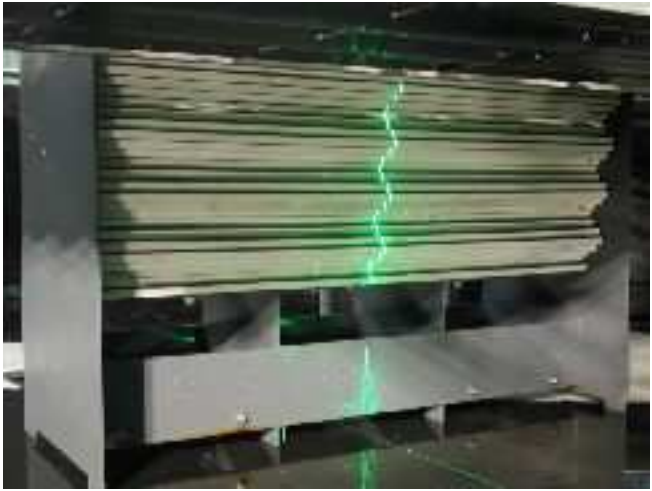}
\caption{Perspective view of the wave generator before the filling
of the tank. The thin vertical line in the middle of the plates is
the impact of the laser sheet, used for the PIV visualization. We
can see the two vertical sidewalls of the generator that hinder to
view the field very close to the generator.} \label{fig:photo}
\end{figure}

The generator consists of a pile of 24 expanded PVC sheets
($2\times36\times150$ cm) enclosed in a parallelepiped half-opened
box and free to slip one over the other. The plates are weighted
with thin lead sheets so that they are buoyantly neutral in water;
it minimizes friction forces between plates. They are thus separated
by 2.2 mm one from the other. Two rectangular holes in each plates
allow two identical camshafts (see Fig.~\ref{fig:schem_excit}(a)) to
go through the pile, imposing the relative position of the plates.
At rest, the plates are sinusoidally shifted, due to the helicoidal
repartition of the cams (see Fig.~\ref{fig:schem_excit}(b)). The
rotation of the camshafts applies a periodic motion to the plates
which propagates upward (resp. downward) for a clockwise (resp.
anti-clockwise) rotation. The eccentricity of the camshafts defines
the amplitude of oscillation of the plates, namely 6 cm peak to
peak. A perspective view of the generator can be seen in
Fig.~\ref{fig:photo}.

To optimize the emission process, the generator is slightly tilted
in direction of the emitted beam, i.e. the direction of shear
propagation. In the following, we generate a downward propagating
beam (defined by the direction of the group velocity) corresponding
to an upward propagating shear (direction of the phase velocity). As
anticipated, the device can easily generate upward propagating beams
by inverting the sign of rotation of both camshaft.

\subsection{Flow visualization}

We performed experiments in the 13 meters diameter circular tank of
the Coriolis  Plateform in Grenoble, filled with 90 cm of linearly
stratified fluid at 3\% corresponding to a Brunt-V\"ais\"al\"a
period of $T_{BV}=11.7\pm0.2\,\mbox{sec}$. A vertical laser sheet
illuminates the wavemaker (visible in Fig.~\ref{fig:photo}) and the
400 microns diameter particles polystyrene beads seeded in the
stratified fluid. PIV measurements of the velocity field were
performed in order to characterize the waves generated by the new
device. All experiments discussed in this paper were carried out
with a period of excitation $T=36$~s. It corresponds to an angle of
emission $\theta=19^\circ$. Snapshots of the flow every 600~ms were
recorded using a 12bits $1024\times1024$ pixels camera.
Subsequently, we  compiled the CIV correlation algorithms (Fincham
and Delerce 2000) 
 between successive frames to get the
velocity field induced by the wavemaker. Results are reported and
analyzed in the next section.

\section{Analysis of the emanating internal waves}
\label{analysisofinternalwaves}

\subsection{Qualitative analysis}

\begin{figure}[!h]
\includegraphics[width=0.5\columnwidth]{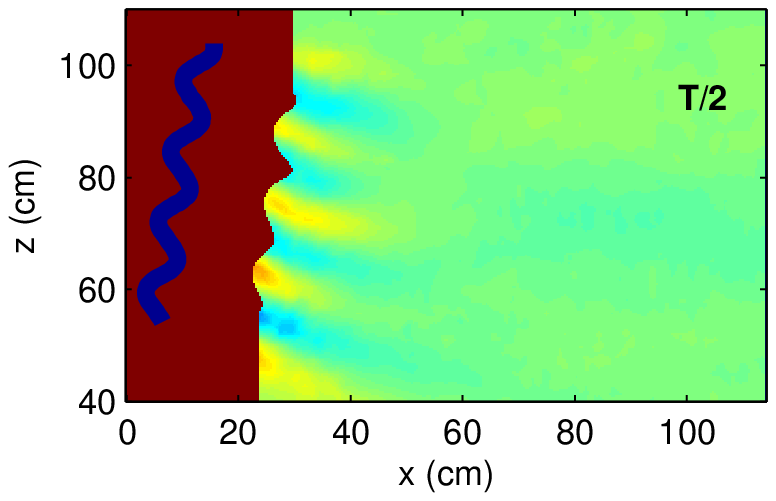}\includegraphics[width=0.5\columnwidth]{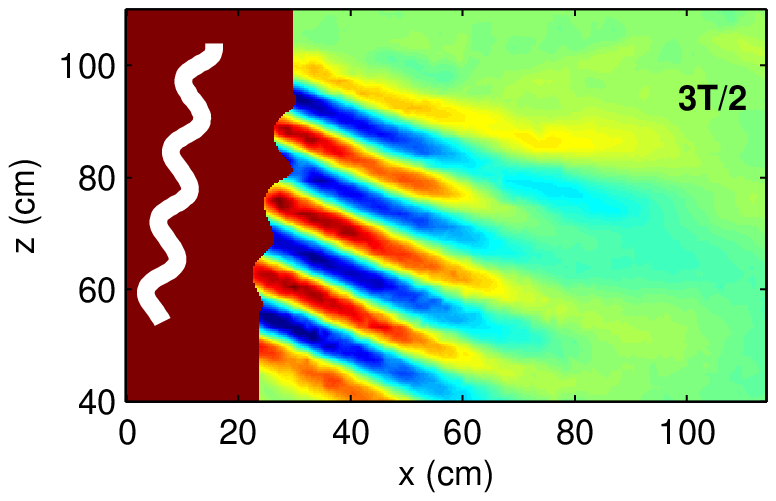}
\includegraphics[width=0.5\columnwidth]{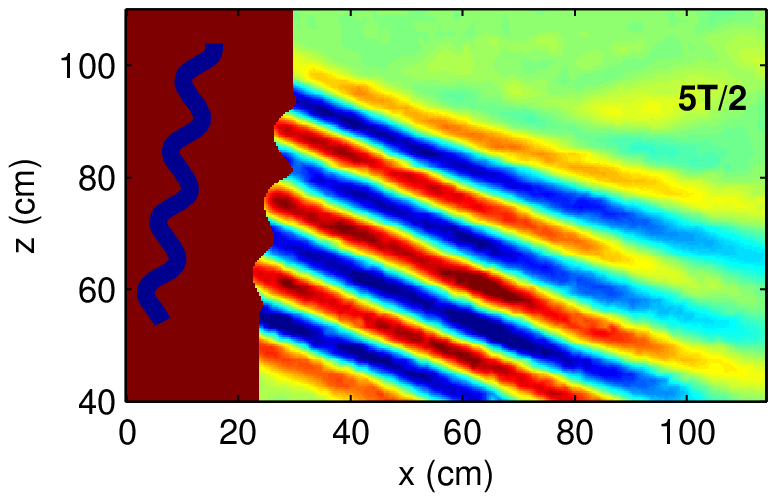}\includegraphics[width=0.5\columnwidth]{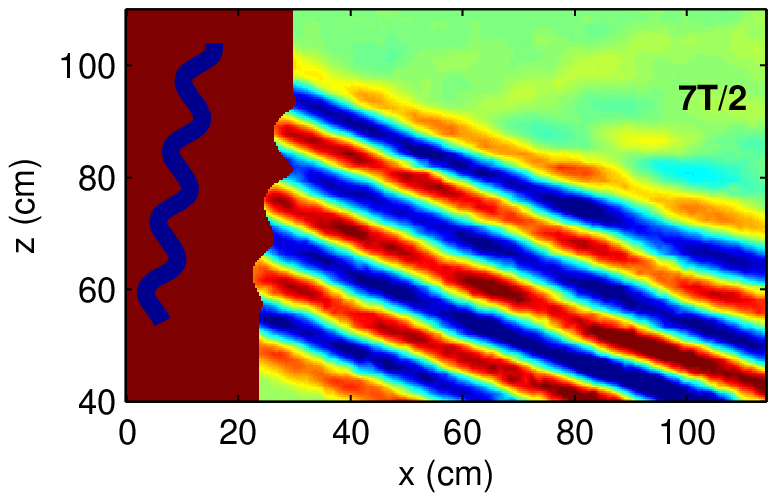}
\caption{(Color online) False-color horizontal velocity pattern in
the vertical plane after 1, 2, 3 and 4 periods $T$ of excitation.
The position of the wavemaker is indicated by the oscillating and
tilted white line. On the left of each panel, the shadowed region
corresponds to the invisible region hidden by the sidewalls of the
generator.} \label{fig:emission}
\end{figure}

Figures~\ref{fig:emission} present four successive images of the
emitted wave. Colors show the horizontal velocity which is the
strongest component in this configuration since the angle of
propagation is rather small. As can be seen, only one downward
propagating beam is emitted: this is the first important improvement
of this new generator. These four pictures clearly emphasize the
generation of the internal plane waves propagating towards the fluid
interior.

The  transversal wave structure perfectly fits also the profile of
the wavemaker, with 4 wavelengths emitted. It is the largest beam
ever generated in a laboratory experiment. It is however important
to stress that there is no theoretical limitation to have a larger
beam. This is of crucial importance for experiments designed to test
plane waves properties.

\begin{figure}[!h]
\includegraphics[width=\columnwidth]{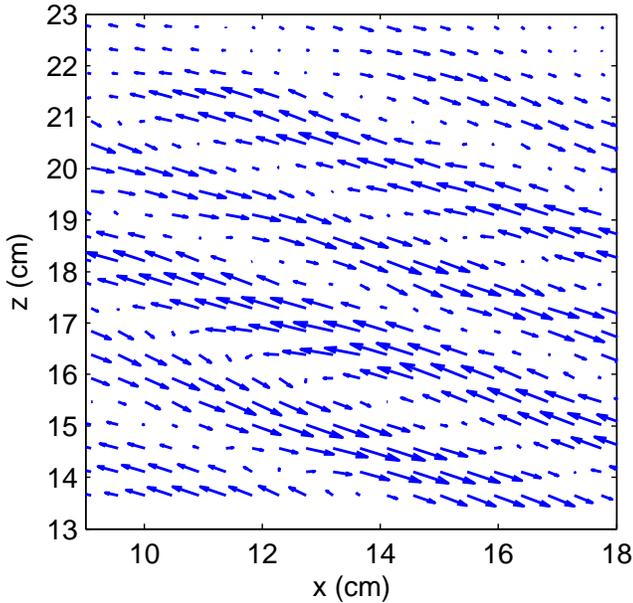}
\caption{Zoom of the velocity field generated by the generator.}
\label{fig:quiver}
\end{figure}

Figure \ref{fig:quiver} shows the velocity field. As expected, it is
parallel to the direction of emission, emphasizing nicely the pure
shear structure. Moreover, it clearly attests that iso-phase lines
are propagating perpendicularly to the energy propagation. The
maximal velocity value measured in that case is $1.8$ mm/s, while
amplitude displacements are of the order of 1 cm.

\subsection{Temporal chromaticity}

\begin{figure}[hbt]
\includegraphics[width=\columnwidth]{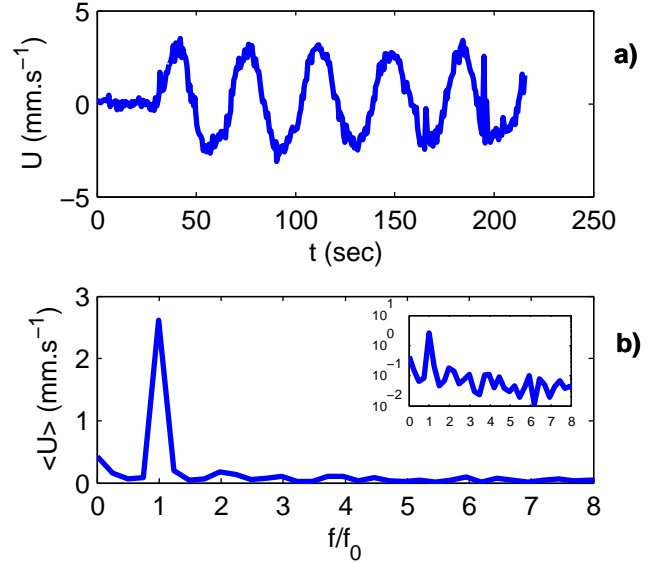}
\caption{Temporal analysis of the internal waves beam in a point
located at 29 cm from the wavemaker and in the middle of the beam.
Panel (a) presents the time evolution of the horizontal velocity
field, while panel (b) shows the Fourier transform of this signal.
Frequencies are renormalized with the excitation one, $f_0=1/T$. The
inset in panel (b) presents the spectrum in logarithmic scale.}
\label{fig:temporal}
\end{figure}

As stated in Sec.~\ref{intro}, {\em monochromatic} internal gravity
waves are highly desirable for studying, for example, the role of
nonlinearities involved in collisions with topography (Ivey and
Nokes 1989, Dauxois and Young 1999, Gostiaux
2006) 
 or with other internal
waves (Teoh et al. 1997)
: the chromaticity quality of the incident beam is therefore of
primary importance.  With the present device, as the forcing is
itself solution of the wave equation, no harmonics are measurable in
the flow. To attest this result, Fig.~\ref{fig:temporal}(a) shows a
temporal evolution of the horizontal velocity at a point situated in
the centerline of the beam, 29 cm away from the generator.

First, note that the signal is vanishingly small before starting to
increase at $t=30$ s. This remark allows to compute the group
velocity of the beam. As the generation and measurement points are
distant by 29 cm, one gets a group velocity $c_g\simeq0.97$ cm/s.
This value has to be compared with the theoretical one,
$c_g=\lambda/(T\tan\theta)$, where $T$ is the excitation period,
$\lambda$ the wavelength of the internal beam and $\theta$ the angle
of propagation. With $\lambda=13.2$ cm, $T=36$ s and
$\theta=19^\circ$, the expected value $c_g=$ 1.06cm/s compares well
with the measurement.

Moreover, one can notice that the transient regime is almost
invisible as the wave reaches the measurement area. This allows to
consider the forcing of such a wave as a suddenly switched on
sinusoid (sinus function convoluted with a Heaviside distribution),
which allows fine comparisons with analytical results using such a
forcing (Dauxois and Young 1999).

Once the beam reaches the measurement point, the velocity oscillates
at a well defined frequency. The temporal spectrum of the wave,
plotted in Fig.~\ref{fig:temporal}(b), confirms that the frequency
signal is, as expected, imposed by the generator. Moreover, it shows
that amplitudes of higher harmonics (see the inset in logarithmic
scale) are at least one order of magnitude smaller than the
fundamental one. The emitted beam is thus highly temporally
monochromatic. The latter characteristic allows to filter the
velocity field at the excitation frequency in order to reduce
measurement noise. The filtering technique is used in
Figs.~\ref{fig:spatial} and~\ref{fig:damping} which present spatial
properties of the beam.

\subsection{Spatial  structure of the beam}

\begin{figure}[ht]
\includegraphics[width=\columnwidth]{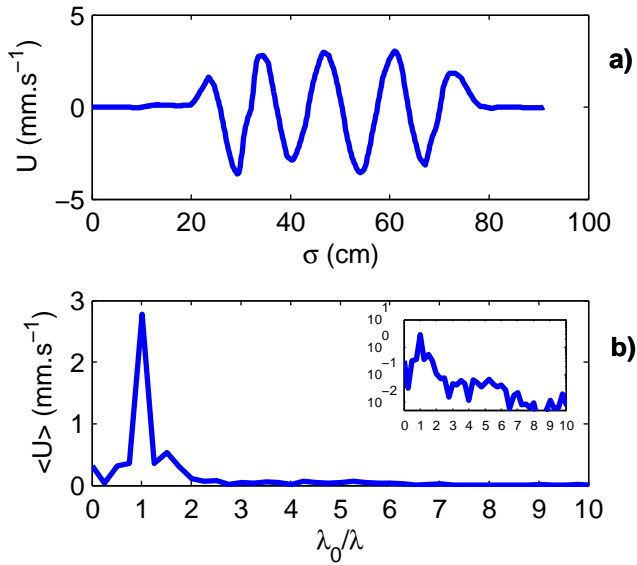}
\caption{Transversal structure of the internal waves beam, 29 cm
from the wavemaker. Panel (a) presents the spatial evolution of the
horizontal velocity field, while panel (b) shows the Fourier
transform of this signal. Wavelengths are renormalized with the
excitation one $\lambda_0$. The inset in panel (b) presents the
spectrum in logarithmic scale.}\label{fig:spatial}
\end{figure}

Contrary to optical or acoustic waves, and owing to the unusual
dispersion relation~(\ref{disprelat}) of internal waves, temporal
monochromatism does not imply any spatial monochromatism: it is
necessary to study them separately. Fig.~\ref{fig:spatial}(a)
presents a cross section of the horizontal velocity field at 29~cm
from the wavemaker. One may easily distinguish slightly more than
four wavelengths. Its corresponding wavelength spectrum is shown in
Fig.~\ref{fig:spatial}(b). The spatial monochromatism of the
produced beam is attested by the very well defined peak of the
spectrum. Note that a careful study shows that end effects in the
horizontal $y$-direction have very little, if no, consequences: one
can only detect a very small transverse component of the velocity
field.

\begin{figure}[ht]
\includegraphics[width=\columnwidth]{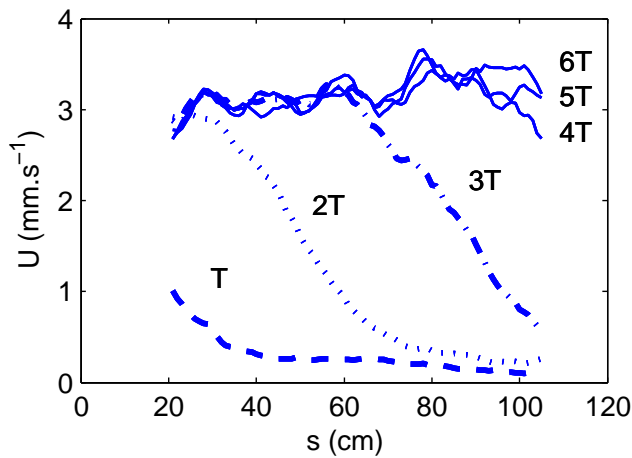}
\caption{Superposition of successive longitudinal profiles of the
horizontal velocity after 1, 2, 3, 4, 5 and 6 excitation periods.}
\label{fig:damping}
\end{figure}

Finally, Fig.~\ref{fig:damping} emphasizes that the viscous damping
of the beam along its propagation is extremely weak. One can notice
the non-dispersive profile of the wave front: the beam does not
disperse since it is spatially monochromatic. From this picture, it
is possible to get another reliable estimate of the group velocity.
Indeed, by measuring the location difference of the appearance of
the velocity $U=1$ mm/s or by trying to superpose the first three
profiles, one gets $c_g\simeq 38$ cm per period i. e. 1.06 cm/s: the
precise theoretical value derived previously.

This generator was used to excite waves up to 1.5 meters away from
the region of interest, no appreciable damping was measured at this
point. The absence of nonlinear losses may explain this remarkable
constancy of the wave amplitude.

\vfill\eject
\section{Conclusion}
\label{conclusion}

We have reported the design of a new kind of internal waves
generator that imposes a rigid  boundary condition compatible with
the equations governing the propagation of plane internal waves. By
generating a propagating shear, we select a single direction of
emission of the waves. The waves emitted are purely monochromatic,
both spatially and temporally, which is of first importance for the
study of nonlinear processes involved in internal waves dynamics.

Finally, it is important to emphasize that this experimental setup
allows to generate any kind of wave form. Not only sinusoidal, but
also gaussian beams or more complicated shapes can be easily
generated, by modifying the camshafts eccentricity appropriately.
This possibility opens new perspective for future studies of the
dynamics of internal waves in stratified fluids.

\begin{acknowledgements}
We warmly thank J.~Sommeria for helpful discussions and S.~Viboud
for help during the experiments. Figure~\ref{fig:schem_emission}(f)
was obtained by D. Benielli and J. Sommeria. Comments to the
manuscript by Denis Martinand are deeply appreciated. This work has
been partially supported by 2005-ANR Project TOPOGI-3D and by the
2006-PATOM CNRS program.
\end{acknowledgements}


\bibliographystyle{spbasic}      


\end{document}